\documentclass[12pt]{article}
\usepackage{scicite}
\usepackage{times}
\usepackage{graphicx}
\usepackage{helvet}
\newenvironment{sciabstract}{%
\begin{quote} \bf}
{\end{quote}}

\newcounter{lastnote}
\newenvironment{scilastnote}{%
\setcounter{lastnote}{\value{enumiv}}%
\addtocounter{lastnote}{+1}%
\begin{list}%
{\arabic{lastnote}.} {\setlength{\leftmargin}{.22in}}
{\setlength{\labelsep}{.5em}}} {\end{list}}

\title{Giant Fluctuations of Coulomb Drag in a Bilayer System}

\author
{A. S. Price,$^{1}$ A. K. Savchenko,$^{1\ast}$ B. N. Narozhny,$^{2}$ G. Allison,$^{1}$ D. A. Ritchie$^{3}$\\
\\
\normalsize{$^{1}$School of Physics, University of Exeter, Stocker
Road,
  Exeter, EX4 4QL, UK}\\
\normalsize{$^{2}$The Abdus Salam ICTP, Strada Costiera 11,
Trieste
  I-34100, Italy}\\
\normalsize{$^{3}$Cavendish Laboratory, University of Cambridge, Madingley Road, Cambridge CB3 0HE, UK}\\
\\
\normalsize{$^\ast$To whom correspondence should be addressed:
a.k.savchenko@ex.ac.uk.} }

\date{}

\begin{document}

\baselineskip24pt

\maketitle

\begin{sciabstract}
  The Coulomb drag in a system of two parallel layers is the
result of electron-electron interactions between the layers. We
have observed reproducible fluctuations of the drag, both as a
function of magnetic field and electron concentration, which are a
manifestation of quantum interference of electrons in the layers.
At low temperatures the fluctuations exceed the average drag,
giving rise to random changes of the sign of the drag. The
fluctuations are found to be much larger than previously expected,
and we propose a model which explains their enhancement by
considering fluctuations of local electron properties.
\end{sciabstract}

In conventional measurements of the resistance of a two-dimensional
(2D) layer an electrical current is driven through the layer and the
voltage drop along the layer is measured. In contrast, Coulomb drag
studies are performed on two closely spaced but electrically
isolated layers, where a current $I_1$ is driven through one of the
layers (active layer) and the voltage drop $V_2$ is measured along
the other (passive) layer (Fig. 1). The origin of this voltage is
electron-electron (\emph{e-e}) interaction between the layers, which
creates a `frictional' force that drags electrons in the second
layer. The ratio of this voltage to the driving current $R_D =
-V_2/I_1$ (the drag resistance) is a measure of \emph{e-e}
interaction between the layers. The measurement of Coulomb drag in
systems of parallel layers was first proposed in Ref.
\cite{Pogrebinskii77, Price83} and later realised in a number of
experiments \cite{SolomonPRL89,GramilaPRL91,SivanPRL92,Lilly98,Lok}
(for a review see Ref. \cite{Rojo99}). As Coulomb drag originates
from \emph{e-e} interactions, it has become a sensitive tool for
their study in many problems of contemporary condensed matter
physics. For example, Coulomb drag has been used in the search for
Bose-condensation of interlayer excitons \cite{SnokeScience02}, the
metal-insulator transition in two-dimensional (2D) layers
\cite{PillarisettyPRB05}, and Wigner crystal formation in quantum
wires \cite{YamamotoScience06}.

Electron-electron scattering, and the resulting momentum transfer
between the layers, usually creates a so-called `positive' Coulomb
drag, where electrons moving in the active layer drag electrons in
the passive layer in the same direction. There are also some cases
where unusual, `negative' Coulomb drag is observed: e.g. between 2D
layers in the presence of a strong, quantising magnetic field
\cite{Lilly98, Lok}; and between two dilute, one-dimensional wires
where electrons are arranged into a Wigner crystal
\cite{YamamotoScience06}. All previous studies of the Coulomb drag,
however, refer to the macroscopic (average) drag resistance.
Recently there have been theoretical predictions of the possibility
to observe random fluctuations of the Coulomb drag
\cite{NarozhnyPRL00,MortensonPRB02}, where the sign of the
frictional force will change randomly from positive to negative when
either the carrier concentration, $n$, or applied (very small)
magnetic field, $B$, are varied.

Drag fluctuations originate from the wave nature of electrons and
the presence of disorder (impurities) in the layers. Electrons
travel around each layer and interfere with each other, after
collisions with impurities, over the characteristic area $\sim
L_\varphi^2$, where $L_\varphi$ is the coherence length (Fig. 1).
This interference is very important for conductive properties of
electron waves. For example, the interference pattern is changed
when the phase of electron waves is varied by a small magnetic
field, producing universal conductance fluctuations (UCF) seen in
small samples with size $L\sim L_\varphi$. There is, however, a
significant difference between UCF and the fluctuations of the drag
resistance. The former are only a small correction to the average
value of the conductance: in our experiment the single-layer
resistance fluctuates by $\sim200$ mOhm around an average resistance
of approximately 500 Ohm. In contrast, the drag fluctuations,
although small in absolute magnitude ($\sim 20$ mOhm) are able to
change randomly, but reproducibly the sign of the Coulomb drag
between positive and negative. Surprisingly, we have found that
these fluctuations of the Coulomb drag, observed at temperatures
below 1 K, are four orders of magnitude larger than predicted in
Ref. \cite{NarozhnyPRL00}.

Our explanation of the giant drag fluctuations takes into account
that, unlike the UCF, the drag fluctuations are not only an
interference but also fundamentally an interaction effect. In
conventional drag structures the electron mean free path $l$ is much
larger than the separation $d$ between the layers, and therefore
large momentum transfers $\hbar q$ between electrons in the layers
become essential. According to the quantum mechanical uncertainty
principle, $\Delta r \Delta q \sim 1$, electrons interact over small
distances $\Delta r\ll l$ when exchanging large values of momentum
(Fig. 1). As a result the local properties of the layers, such as
the local density of electron states (LDoS), become important in the
interlayer \emph{e-e} interaction. These local properties at the
scale $\Delta r \ll l$ exhibit strong fluctuations \cite{lerner}
that directly manifest themselves in the fluctuations of the Coulomb
drag.

The samples used in this work are AlGaAs-GaAs double-layer
structures, in which the carrier concentration of each layer can be
independently controlled by gate voltage. The two GaAs quantum wells
of the structure, $200\,${\AA} in thickness, are separated by an
$\mathrm{Al_{0.33}Ga_{0.67}As}$ layer of thickness $300\,${\AA}.
Each layer has a Hall-bar geometry, $60\,\mathrm{\mu m}$ in width
and with a distance between the voltage probes of $60\,\mathrm{\mu
m}$ \cite{som}.

Figure 2 shows the appearance of the fluctuations in the drag
resistivity, $\rho_D$, at low temperatures. At higher temperatures,
the drag resistance changes monotonically with both $T$ and $n$: the
insets to Fig. 2 show that $\rho_D$ increases with increasing
temperature as $T^2$ and decreases with increasing passive-layer
carrier concentration as $n_2^{b}$, where $b \approx -1.5$. These
results are consistent with existing experimental work on the
average Coulomb drag ~\cite{GramilaPRL91,KelloggSSC02}.

Figure 3A shows a zoomed-in view of the reproducible fluctuations as
a function of $n_2$. These fluctuations result in an alternating
sign of the drag, which is demonstrated in the inset to Fig. 3 where
the temperature dependence of the drag is shown at two different
values of $n_2$. The drag is seen first to decrease as the
temperature is decreased, but then become either increasingly
positive or increasingly negative, dependent upon $n_2$. The
reproducible fluctuations of the drag resistivity have also been
observed as a function of magnetic field (Fig. 3B). For a fixed
temperature, the magnitude of the drag fluctuations as a function of
$n_2$ is roughly the same as that as a function of $B$.

The theory of Ref. \cite{NarozhnyPRL00} calculates the variance of
drag fluctuations in the so-called diffusive regime, $l<d$. In this
case the drag is determined by global properties of the layers,
averaged over a region $\Delta r \gg l$. The expected variance of
drag fluctuations (at low $T$ when the fluctuations exceed the
average) in the diffusive regime is
\begin{equation}
\label{eq:dragvariance}%
\langle\Delta\sigma_D^2\rangle \approx A
\frac{e^4}{\hbar^2}\frac{E_T(L) \tau_\varphi \ln{\kappa
d}}{g^4\hbar(\kappa d)^3},
\end{equation}
where $\sigma_D \approx \rho_D / (\rho_1 \rho_2)$, and $\rho_1$ and
$\rho_2$ are the active and passive layer resistivities,
respectively; $E_T(L)$ is the Thouless energy, $E_T(L) = \hbar D /
L^2$, $D$ is the diffusion coefficient; $\tau_\varphi$ is the
decoherence time; $\kappa$ is the inverse screening length; $A =
4.9\times 10^{-3}$ and $g = h/(e^2\rho)$ is the dimensionless
conductivity of the layers. Using the parameters of our system, this
expression gives a variance of $\sim 6\times 10^{-11}\,\mathrm{\mu
S^2}$, which is approximately eight orders of magnitude smaller than
the variance of the observed drag fluctuations. The fluctuations in
$\rho_D$ have been measured in two different samples, and their
variance is seen to be similar in magnitude and $T$-dependence,
confirming the discrepancy with the theoretical prediction
\cite{NarozhnyPRL00}.

The expected fluctuations of the drag conductivity share the same
origin as the UCF in the conventional conductivity: coherent
electron transport over $L_\varphi$ in the layers prior to
\emph{e-e} interaction between the layers (Fig. 1). For this reason
we have compared the drag fluctuations with the fluctuations seen in
the single-layer resistivity of the same structure (Fig. 3B, inset),
which have shown the usual behaviour \cite{ucf}. We estimate the
expected variance of the single-layer conductance fluctuations using
the relation $\label{eq:slvariance}\langle \Delta \sigma_{xx}^2
\rangle = \left({e^2}/{h}\right)^2\left({L_T}/{L}\right)^2$, where
$L_T = \sqrt{\hbar D /k_B T}$ is the thermal length \cite{ucf}. This
expression produces a value of $0.8\,\mathrm{\mu S^2}$, which is in
good agreement with the measured value of $0.6\,\mathrm{\mu
  S^2}$. The typical `period' of the drag fluctuations (the
correlation field, $\Delta B_c$) is similar to that of the UCF
\cite{som}, indicating that both depend upon the same $L_\varphi$
and have the same quantum origin.

To address the question of the discrepancy between the magnitude of
drag fluctuations in theory \cite{NarozhnyPRL00} and our
observations, we stress that the theoretical prediction for the
variance, Eq.~\ref{eq:dragvariance}, was obtained under the
assumption of diffusive motion of interacting electrons, with small
interlayer momentum transfers, $q\ll1/l$. As the layers are
separated by a distance $d$, the \emph{e-e} interactions are
screened at distances $\Delta r > d$. Therefore, in all regimes the
maximum momentum transfers are limited by $q<1/d$. In the diffusive
regime, $l<d$, this relation also means that $q<1/l$, that is,
interlayer \emph{e-e} interactions occur at distances $\Delta r > l$
and involve scattering by many impurities in the individual layers.
In the opposite situation, $l\gg d$, the transferred momenta will
include both small and large $q$-values: $q<1/l$ and $1/l<q<1/d$. We
have seen that small $q$ cannot explain the large fluctuations of
the drag \cite{NarozhnyPRL00}, and so argue that it is large
momentum transfers with $q>1/l$ which give rise to the observed
effect. In this case the two electrons interact at a distance
$\Delta r$ that is smaller than the average impurity separation and,
therefore, it is the local electron properties of the layers which
determine \emph{e-e} interaction. In Ref.\cite{lerner} it is shown
that the fluctuations of the local properties are larger compared to
those of the global properties that are responsible for the drag in
the diffusive case.

A theoretical expression for the drag conductivity is obtained by
means of a Kubo formula analysis
\cite{KamenevPRB95,ZhengPRB93,Jauho93,FlensbergPRB95} (detailed
description in supporting text). For a qualitative estimate, three
factors have to be taken into account: (i) the inter-layer matrix
elements of the Coulomb interaction ${\cal D}_{ij}$; (ii) the phase
space (the number of electron states available for scattering); and
(iii) the electron-hole ({\it e-h}) asymmetry in both layers. Point
(iii) takes into account that in a quantum system the current is
carried by both electron-like (above the Fermi surface) and
hole-like (below the Fermi surface) excitations. If they were
completely symmetric with respect to each other, then the
current-carrying state of the active layer would have zero total
momentum and thus no drag effect would be possible. The physical
quantity that measures the degree of {\it e-h} asymmetry is the
non-linear susceptibility $\Gamma$ of the 2D layer. Theoretically,
the drag conductivity is represented in terms of the non-linear
susceptibilities of each layer and dynamically screened interlayer
Coulomb interaction ${\cal D}_{ij}(\omega)$ as $\sigma_{D} \propto
\int d\omega{\cal D}_{12}(\omega) \Gamma_{2}(\omega) {\cal
D}_{21}(\omega) \Gamma_{1}(\omega)$ (indices 1 and 2 correspond to
the two layers) \cite{KamenevPRB95,NarozhnyPRL00}. The {\it e-h}
asymmetry appears in $\Gamma$ as a derivative of the density of
states $\nu$ and the diffusion coefficient $D$: $\Gamma\propto
\partial\left(\nu D\right) / \partial \mu$, and it is this
quantity that is responsible for the fact that drag fluctuations
can exceed the average. As $D \nu \sim g$ and the typical energy
of electrons is the Fermi energy, $E_F$, we have ${\partial(\nu
D)}/{\partial \mu} \sim g/E_F$ for the average drag. The typical
energy scale for the interfering electrons, however, is
$E_T(L_\varphi)$ ~\cite{ucf}, which is much smaller than $E_F$ and
therefore a mesoscopic system has larger \emph{e-h} asymmetry.

Under the condition of large momentum transfer between the layers,
fluctuations in $\Gamma$ are similar to the fluctuations of the
LDoS, which can be estimated as $\delta\nu^2
\sim({\nu^2}/{g})\ln{({\rm max}(L_\varphi, L_T)}/{l})$
\cite{lerner}. Also, the interaction in the ballistic regime can be
assumed to be constant, ${\cal D}_{ij} \approx -1/\nu\kappa d$, as
$q$ is limited by $q\leq1/d$. Finally, to average fluctuations of
the drag over the sample with size $L$ we should divide it into
coherent patches of size $L_\varphi$ that fluctuate independently
and thus decrease the total variance: $\langle\Delta\sigma_D^2
\rangle = \langle \Delta\sigma_D^2 (L_\varphi)
\rangle\left({L_\varphi}/{L} \right)^2$. If $k_B T>E_T(L_\varphi)$,
fluctuations are further averaged on the scale of $\sim k_B T$, and
therefore the variance is suppressed by an additional factor of
$E_T(L_\varphi)/k_BT$. Combining the above arguments we find

\begin{equation}
\langle\Delta\sigma_D^2\rangle = N \frac{e^4}{g^2\hbar^2(\kappa
d)^4}\frac{(k_B T)^2}{E_T^2(L_\varphi)}
\frac{l^4L_\varphi^2}{d^4L^2}, \label{BallisticDragVariance}
\end{equation}

\noindent where $N$ is a numerical coefficient.

Compared to the diffusive situation (Eq. \ref{eq:dragvariance})
the fluctuations described by our model are greatly enhanced. The
difference between Eqs. ~\ref{BallisticDragVariance} and
\ref{eq:dragvariance} comes from the fact that in the ballistic
regime electrons are not scattered by impurities between events of
\textit{e-e} scattering. Large momentum transfers correspond to
short distances, and thus in the ballistic regime drag
measurements explore the local (as opposed to averaged over the
whole sample) non-linear susceptibility. This leads to the
appearance of three extra factors in Eq.
~\ref{BallisticDragVariance}: (i) the factor $l^4/d^4$ (which is
also present in the average drag in the ballistic regime -- see
Ref. \cite{KamenevPRB95}); (ii) the phase space factor $T/E_T$
(which appears due to the fact that interaction parameters ${\cal
D}_{ij}$ are now energy-independent); and (iii) the extra factor
$g^2$ due to fluctuations of the local non-linear susceptibility.
Local fluctuations are enhanced since the random quantity $\Gamma$
is now averaged over a small part of the ensemble, allowing one to
detect rare impurity configurations.

Our model not only explains the large magnitude of the fluctuations,
but also predicts a non-trivial temperature dependence of their
magnitude. The latter comes from the change in the temperature
dependence of $L_\varphi$ \cite{NarozhnyPRB02}: at low temperatures,
$k_B T\tau / \hbar\ll 1$, the usual result is $L_\varphi \propto
T^{-1/2}$, while for $k_B T\tau / \hbar >1$ the temperature
dependence changes to $L_\varphi \propto T^{-1}$ \cite{aar}.
Consequently, the temperature dependence of the variance of the drag
fluctuations is expected to change from $T^{-1}$ at low $T$, to
$T^{-4}$ at high $T$. This temperature dependence is very different
from the $T$-dependence of drag fluctuations in the diffusive
regime, $\langle \Delta \sigma_D^2\rangle \propto T^{-1}$. To test
the prediction of Eq.~\ref{BallisticDragVariance}, the
$T$-dependence of $\langle \Delta \sigma_D^2\rangle$ has been
analysed (Fig. 4). The variance is calculated in the limits of both
the diffusive $\tau_\varphi$ (solid line, $\tau_\varphi^{-1} \propto
T$) and ballistic $\tau_\varphi$ (dashed line, $\tau_\varphi^{-1}
\propto T^2$), using $N \simeq 10^{-4}$. In fitting the drag
variance we have found $\tau_\varphi$ to agree with theory to within
a factor of two \cite{som}, which is typical of the agreement found
in other experiments on determining $\tau_\varphi$ \cite{beenakker}.
(The single-layer values of $\tau_\varphi$ found from our analysis
of the UCF agree with theory to within a factor of $1.5$.) Thus, the
temperature dependence of the observed drag fluctuations strongly
supports the validity of our explanation.

We have observed reproducible fluctuations of the Coulomb drag and
demonstrated that they are an informative tool for studying wave
properties of electrons in disordered materials, and the local
properties in particular. Contrary to UCF which originate from
quantum interference, fluctuations of drag result from an
interplay of the interference and \textit{e-e} interactions. More
theoretical and experimental work is required to study their
manifestation in different situations. For instance, similarly to
the previous extensive studies of the evolution of UCF with
increasing magnetic field, such experiments can be performed on
the fluctuations of drag. One of the important developments in the
field of Coulomb drag fluctuations can be their study in
quantising magnetic fields, including the regimes of integer and
fractional quantum Hall effects.

\newpage

\bibliographystyle{Science}

\begin{thebibliography}{10}
\bibitem{som}
The details of the structures, measurements and the model are
available as
  supporting material on Science Online.
\end{thebibliography}

\begin{scilastnote}
\item Authors thank I.L. Aleiner, M. Entin, I.L. Lerner,
A. Kamenev, and A. Stern for numerous helpful discussions.\\

\end{scilastnote}

\noindent
\textbf{Supporting Online Material}\\
www.sciencemag.org\\
Materials and Methods\\
SOM text\\
Figs. S1 to S3\\

\newpage

\includegraphics[width = 12cm]{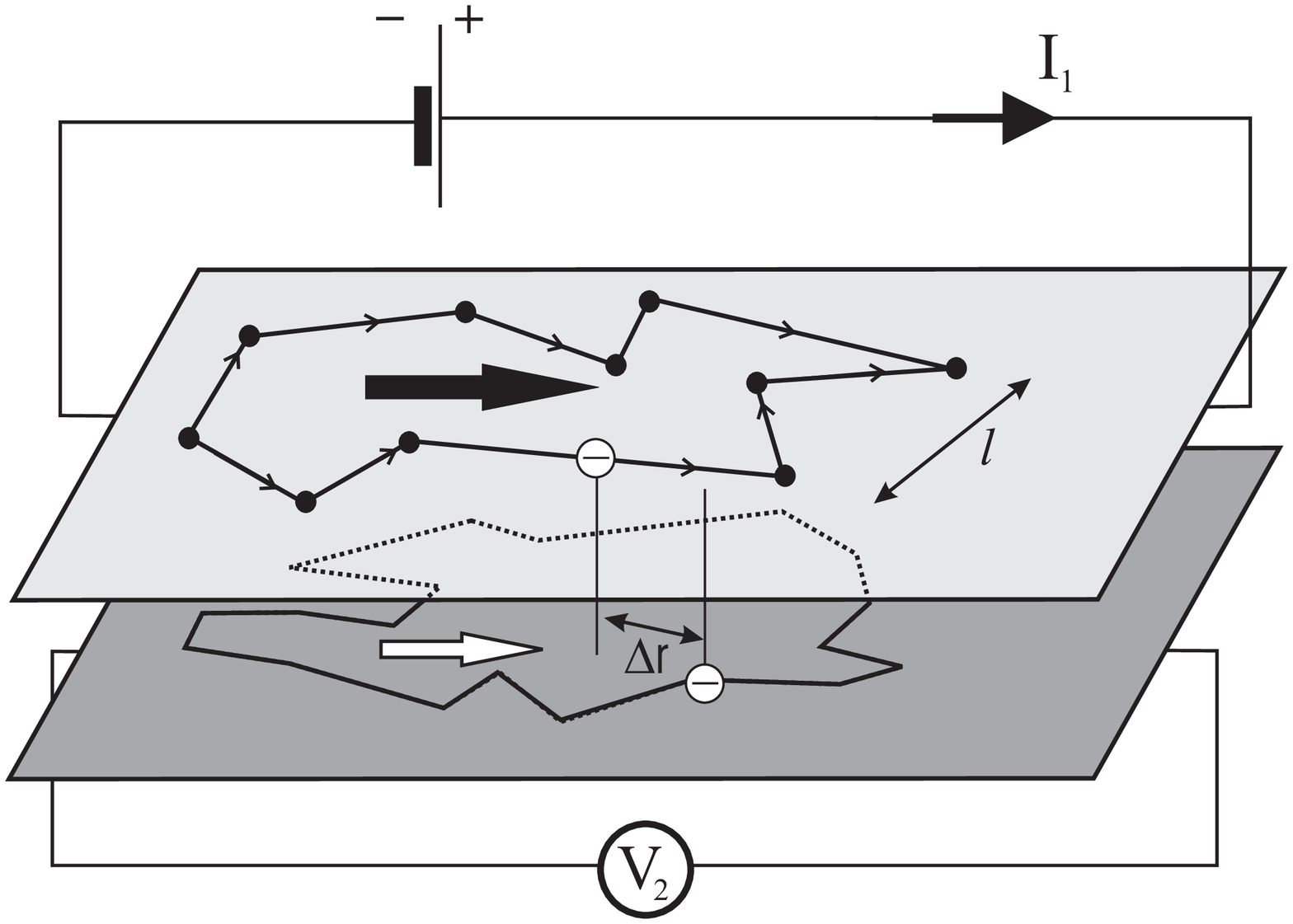}\\
\textsf{\textbf{Fig. 1.} Schematic showing the origin of the drag
signal $\mathsf{V_2}$ induced by the current $\mathsf{I_1}$. The
fluctuations of the drag arise from the interference of electron
waves in each layer, before the two electrons take part in the
interlayer interaction.}
\\
\noindent

\includegraphics[width = 12cm]{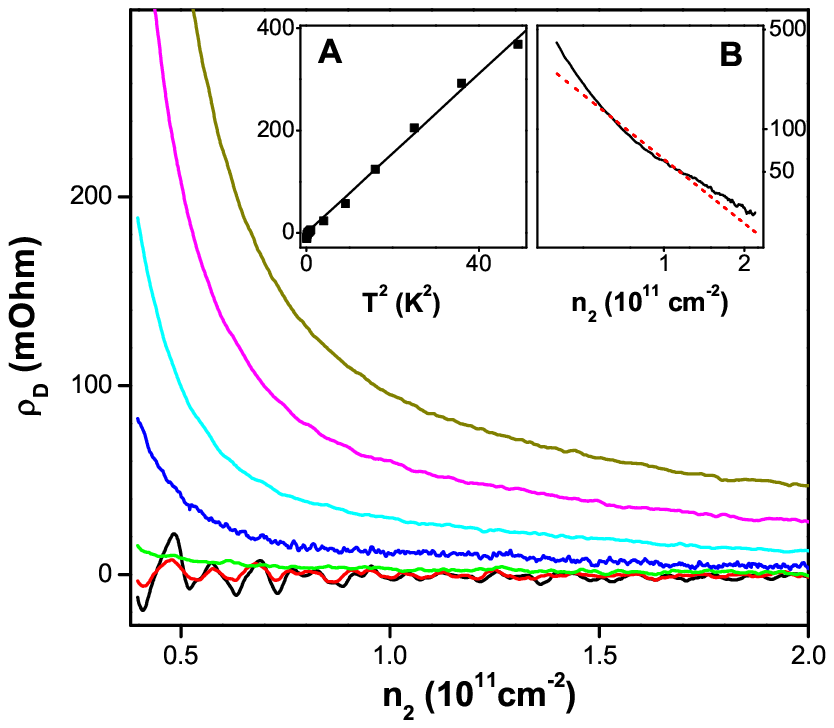}\\
\textsf{\textbf{Fig. 2.} Drag resistivity as a function of
passive-layer carrier concentration for different temperatures:
$\mathsf{T = 5, 4, 3, 2, 1, 0.4}$, and $\mathsf{0.24}$ K, from top
to bottom. Inset (\textbf{A}): $\mathsf{\rho_D}$ as a function of
$\mathsf{T^2}$. Inset (\textbf{B}): $\mathsf{\rho_D}$ as a
function of $\mathsf{n_2}$, with $\mathsf{n_1 = 1.1\times
10^{11}\,cm^{-2}}$; dashed line is a $\mathsf{n_2^{-1.5}}$ fit.}
\\
\noindent

\includegraphics[width = 12cm]{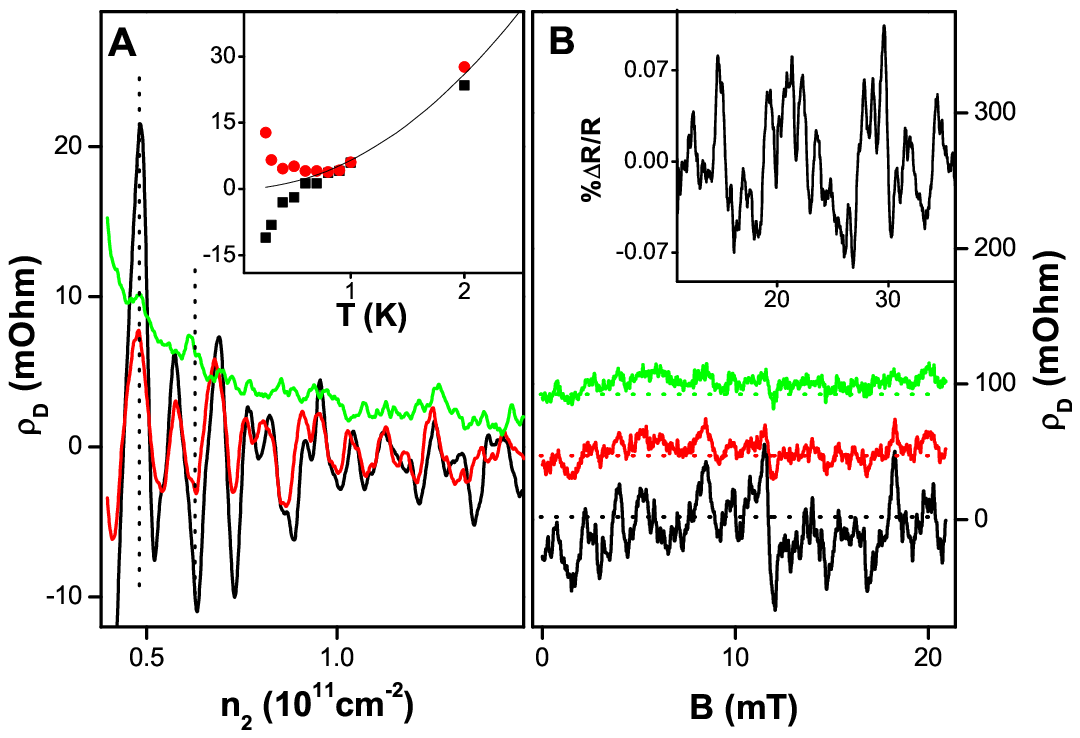}\\
\textsf{\textbf{Fig. 3.} (\textbf{A}) Drag resistance measured at
low temperatures as a function of passive-layer concentration;
$\mathsf{T = 1, 0.4}$, and $\mathsf{0.24}$ K, from top to bottom.
Inset: $\mathsf{\rho_D}$ as a function of $\mathsf{T}$ for two
values of $\mathsf{n_2}$ denoted by the dotted lines in Fig. 3A;
solid line is the expected $\mathsf{T^2}$ dependence of the
average drag. (\textbf{B}): $\mathsf{\rho_D}$ as a function of
$\mathsf{B}$; $\mathsf{T = 0.4, 0.35}$, and $\mathsf{0.24}$ K,
from top to bottom. (Graphs for higher $\mathsf{T}$ are vertically
offset for clarity.) Single-layer concentration for each layer is
$\mathsf{5.8\times 10^{10} cm^{-2}}$. Inset: The UCF of the
single-layer, with an average background resistance of 500 Ohm
subtracted.}
\\
\noindent

\includegraphics[width = 12cm]{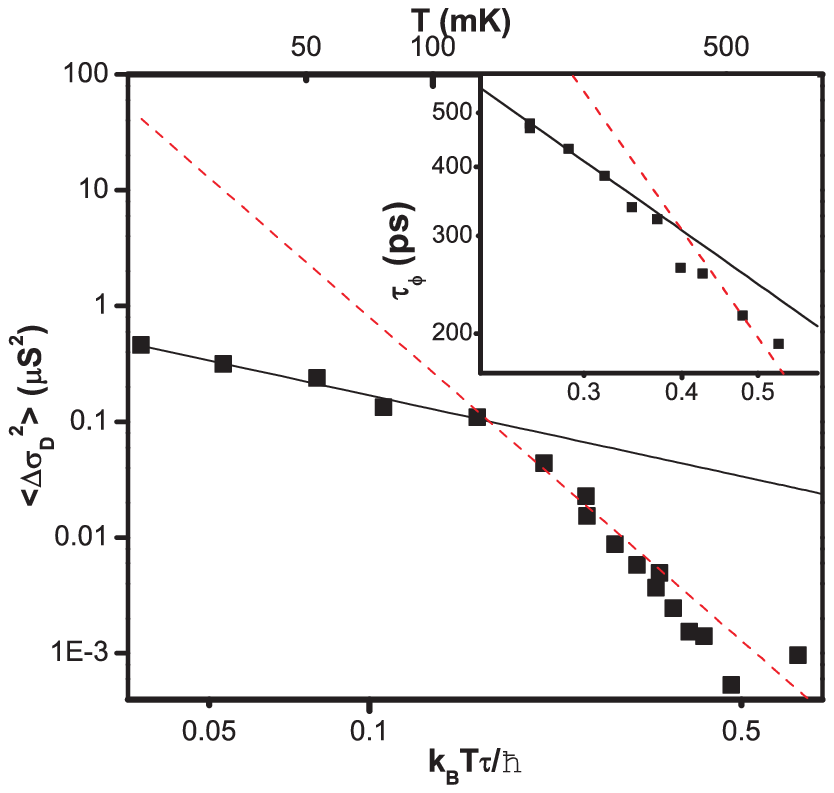}\\
\textsf{\textbf{Fig. 4.} The variance of the drag conductivity
fluctuations (squares) plotted against temperature. The solid and
dashed lines are calculated using Eq.~\ref{BallisticDragVariance}
with the diffusive and ballistic asymptotes of
$\mathsf{\tau_\varphi}$, respectively. Inset:
$\mathrm{\tau_\varphi}$ extracted from the correlation magnetic
field of the single-layer fluctuations, plotted against
temperature.}

\end{document}